# Birefringence induced polarization-independent and nearly all-angle transparency through a metallic film

Dong-Liang Gao[1], Lei Gao[1 (a)] and Cheng-Wei Qiu[2 (b)]

1) Jiangsu Key Laboratory of Thin Films, Department of Physics, Soochow University, Suzhou 215006, China.
2) Department of Electrical and Computer Engineering, National University of Singapore, 4 Engineering Drive 3, Singapore 117576, Singapore.

**Abstract**

We propose an birefringence route to perfect electromagnetic (EM) wave tunneling through a metallic film which relies on homogeneous birefringent coatings with moderate and positive parameters only. EM transparency is achieved in such an birefringent-metal-birefringent (BMB) structure for both polarizations and over nearly all incident angles. The stringent restrictions in conventional dielectric-metal-dielectric media, i.e., dielectrics with extremely negative permittivity, high magnetic field and polarization dependence (only for TE waves), are not required in our method. The criterion for perfect transmission is obtained by analyzing the effective medium theory and EM fields of such a birefringent structure. The solutions hold for lossless and lossy cases in a quite large frequency range.

a) Email: leigao@suda.edu.cn
b) Email: eleqc@nus.edu.sg

EM wave tunneling through metals has always been intriguing and has received great attention recently [1–7]. In two decades ago, high transparency of classic opaque metallic films had been studied by using prism couplers [8]. Recently, an array of subwavelength holes, which can support surface plasmons (SP's), were proposed to get high transmission through metallic films [1, 9–11]. Meanwhile, metallic films with an array of slits have been studied, where high transmission was also demonstrated [12, 13].

More recently, Zhou et al proposed that perfect transmission through negative permittivity isotropic media [5] can be realized by a dielectric-metal-dielectric (ABA) sandwich structure and the effective medium can be modeled by maximizing density of states (DOS) function [14]. Subsequently, Hooper et al. [6] showed that this perfect transmission is true in the field of optical and quantum mechanical system. The perfect transmission in Ref. [5] holds for a wide range of incident angles, which relies on extremely large permittivity of both layers in order to let the solutions to the transmittance have little dependence of incident angles. And such phenomenon is only present for TE polarized light [6]. In this letter, by introducing birefringence into both sandwich layers, nearly all-angle high transmission can be realized for an BMB sandwich structure for both TE and TM polarized waves. Moreover, the material parameters in the metal layer and birefringent layers are moderate compared to those large figures in conventional ABA structures.

Let us consider a metallic layer M of thickness $d_M$ with permittivity $\varepsilon_M$ ($\varepsilon_M < 0$), which is sandwiched by two identical birefringent and homogeneous layers B. We choose the layers to be parallel to the x-y plane with the z-axis normal to the interfaces of the layers. The constitutive tensor of the relative permittivity in birefringent layer B is assumed to be to be non-magnetic, i.e., $\mu_B = \mu_M = 1$.

We adopt transfer matrix method (TMM) [15, 16] to analyze the electric and magnetic field in the BMB sandwich structure,

$$M_j(\Delta z, \omega, \theta) = \begin{pmatrix} \cos(k_{jz}\Delta z) & -\dfrac{i}{P_j}\sin(k_{jz}\Delta z) \\ -iP_j \sin(k_{jz}\Delta z) & \cos(k_{jz}\Delta z) \end{pmatrix} \quad (1)$$

where $k_{jz}$ denotes the z-component of the wave vector $\vec{k}_j$ in the layer j (j=B,M) with $k_{Bz} = \omega/c\sqrt{\varepsilon_{Bz}}\sqrt{1-\sin^2\theta/\varepsilon_B}$ for TE (TM) polarized waves and $k_{Mz} = \omega/c\sqrt{\varepsilon_M}\sqrt{1-\sin^2\theta/\varepsilon_M}$. In addition, we have $P_B = \sqrt{1-\sin^2\theta/\varepsilon_{Bx}}\sqrt{\varepsilon_{Bx}}$ and $P_M = \sqrt{1-\sin^2\theta/\varepsilon_M}\sqrt{\varepsilon_M}$ for TE polarized waves; $P_B = \sqrt{1-\sin^2\theta/\varepsilon_{Bz}}/\sqrt{\varepsilon_{Bz}}$ and $P_M = \sqrt{1-\sin^2\theta/\varepsilon_M}/\sqrt{\varepsilon_M}$ for TM polarization.

The transmission coefficient of the plane wave can be obtained,

$$t = \frac{2q_0}{(q_0 x_{22} + q_s x_{11}) + (q_0 q_s x_{12} + x_{21})} \tag{2}$$

where $q_0 = q_s = \frac{\sqrt{\varepsilon_0}}{\sqrt{\mu_0}} \sqrt{1 - \frac{\sin^2\theta}{\varepsilon_0 \mu_0}} = \cos\theta$ holds for the vacuum regions z<0 and z>$d_B$ + $d_M$ + $d_B$ ($\varepsilon_0$ = $\mu_0$ = 1), $x_{ij}$ (i, j = 1, 2) are the matrix elements of $X_{BMB}(\omega)$ =$M_B(\omega)M_M(\omega)M_B(\omega)$, representing the total transfer matrix for the proposed BMB sandwich structure.

Similarly to the isotropic ABA cases [5], we consider lossless BMB structures first. The perfect transmission appears when T = $|t|^2$ = 1, from which we can get the following criterion for perfect transmission,

$$\frac{2(P_B - \frac{\cos^2\theta}{P_B})}{\tan(k_{Mz}d_M)} - \frac{(\frac{\cos^2\theta}{P_M} - P_M)}{\tan(k_{Bz}d_B)} - (\frac{P_B^2}{P_M} - \frac{P_M}{P_B^2}\cos^2\theta)\tan(k_{Bz}d_B) = 0 \tag{3}$$

Under the long-wavelength limit ($k_{jz}d_j \to 0$), the whole BMB structure is equivalent to an effective layer of the thickness 2$d_B$+$d_M$ with the effective permittivity tensor $\vec{\varepsilon} = \varepsilon_{Bx} \cdot e_x e_x + \varepsilon_{Bx} \cdot e_y e_y + \varepsilon_{Bz} \cdot e_z e_z$ [17]. For given $\varepsilon_M$, $d_B$ and $d_M$, one can thus choose different $\varepsilon_{Bx}$ and $\varepsilon_{Bz}$ to satisfy $\varepsilon_x^{eff}$ = 1 and $\varepsilon_z^{eff}$ = 1 simultaneously, namely, the effective layer would become transparent with unit transmittance independent of the polarization and the incident angle in principle. Note that $\varepsilon_x^{eff}$ and $\varepsilon_z^{eff}$ cannot be unit at the same time with the choice of $\varepsilon_A$ in isotropic layer A. For instance, if $\varepsilon_x^{eff}$ is adjusted to be unit, the isotropic ABA structure would be transparent only for TE polarized waves.

In contrast, in our letter, we have tan ($k_{zj}d_j$) ~ $k_{zj}d_j$ due to $k_{jz}d_j \to 0$, the third term in Eq. (3) can be ignored. Thus Eq. (3) for TE polarization is reduced to $\varepsilon_x^{eff} = (\varepsilon_{Bx} \cdot 2d_B + \varepsilon_M \cdot d_M)/(2d_B + d_M) = 1$, while Eq. (3) for TM polarized case becomes $\varepsilon_z^{eff} = \frac{\varepsilon_{Bz}\varepsilon_M(2d_B + d_M)}{\varepsilon_{Bz}d_M + \varepsilon_M \cdot 2d_B} = 1$. It is obviously possible to achieve $\varepsilon_x^{eff} = \varepsilon_z^{eff}$ = 1. For large $k_{jz}d_j$, effective medium theory (EMT) is not valid, and one can resort to Eq. (3) directly to find the solutions of $\varepsilon_{Bx}$ and $\varepsilon_{Bz}$ for unit transmittance at given $\varepsilon_M$ and $d_j$. In numerical calculations, we set the thickness of the metallic layer and the birefringent layers to be equal, i.e., $d_B$ = $d_M$ = d.

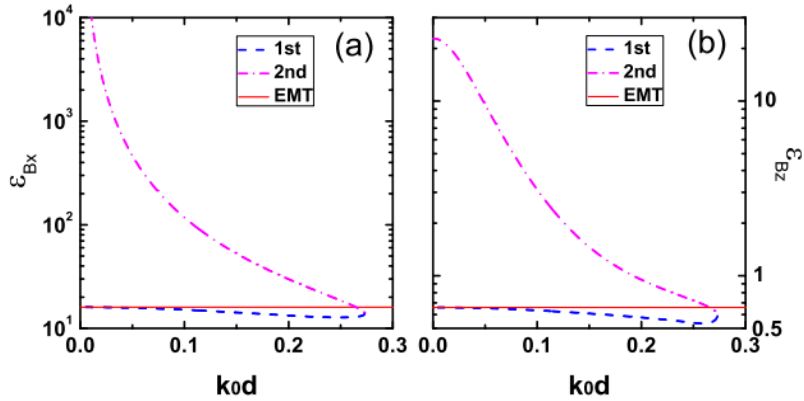

FIG. 1: (Color online) Solutions of Eq. (3) and EMT as a function of the thickness d. Here $\varepsilon_M = -29$ and $\mu_M = 1$; the wavelength of the incident plane wave is $\lambda_0 = 826$nm.

Figure 1 demonstrates the solutions to Eq. (3) as well as EMT for $\varepsilon_{Bx}$ and $\varepsilon_{Bz}$ as a function of the thickness d. We can see that both $\varepsilon_{Bx}$ and $\varepsilon_{Bz}$ have two solutions for a given value of d and the first solutions are close to EMT solutions in the long-wavelength limit. Note that Eq. (3) has solutions even when $k_j d_j$ goes beyond the long-wavelength limit. Taking the first solutions for example, when d=30 nm, we can obtain $\varepsilon_{Bx} = 12.90$ and $\varepsilon_{Bz} = 0.56$. Correspondingly, $k_0 d$ is 0.228 and $k_B d$ is 0.819, which is beyond the long-wavelength limit. Obviously, EMT is no longer accurate under such circumstances.

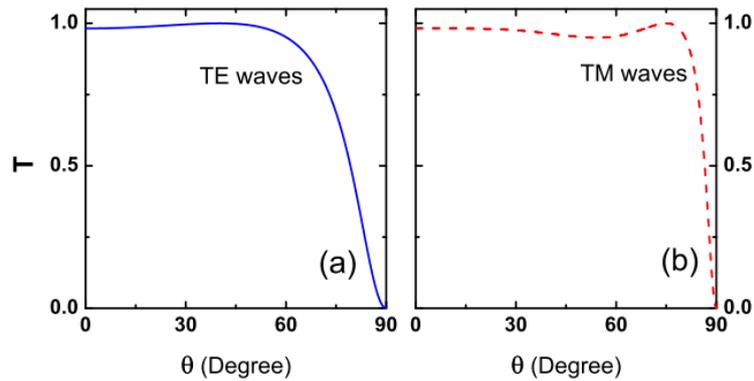

FIG. 2: (Color online) The transmittance T for TE and TM polarized waves against the incident angle $\theta$. The thickness $d_B = d_M = d = 30$ nm, and the other parameters are the same as those in Fig. 1.

As is shown in Fig. (2), the solutions of Eq. (3) for high transmission are almost independent upon the angle and polarization of the incidence. Compared to the isotropic ABA structure [5], our birefringent BMB structure can also achieve nearly all-angle wave tunneling, but different mechanisms are involved. In Ref. [5], the independence of the high transmission upon the incident angle has to rely on the fact that the permittivities of those layers have to be extremely large such that the solutions to the equation governing the high transmission has little variation on the angle θ. In contrast, our BMB recipe simply removes such constraints by employing the birefringence and all parameters involved are realizable. What's more important is that high transmission in both parallel and perpendicular components (with respect to the propagating direction) are obtained, which thus enables the birefringent BMB structure to be simultaneously transparency for TE and TM polarized waves. The above birefringent structure can be realized by periodically stacking two kinds of realistic metallic structures [18].

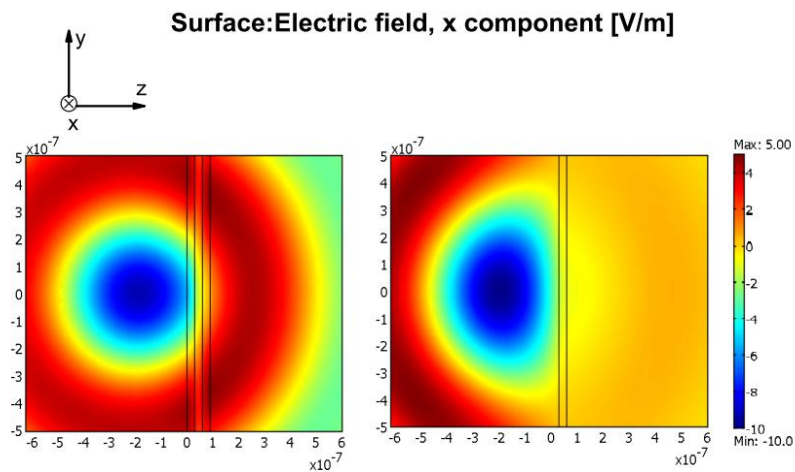

FIG. 3: (Color online) The electric field distribution (a) with the BMB structure and (b) without the BMB structure. The parameters are the same as those in Fig. 1.

In Fig. 3(a), a point source generating TE polarized waves is placed 200 nm away from the surface of the proposed BMB structure. TE polarized waves transmit through the structure in almost all directions without changing the wave fronts. In contrast, a single metallic film without any coatings is far beyond being transparent (see Fig. 3(b)). The numerical simulations for TM polarized point source are similar to the TE ones. In addition, more simulations with Gaussian beams for TE and TM polarization, though suppressed, also confirm the observed features of polarization independence and nearly all-angle transparency in such birefringent BMB structures.

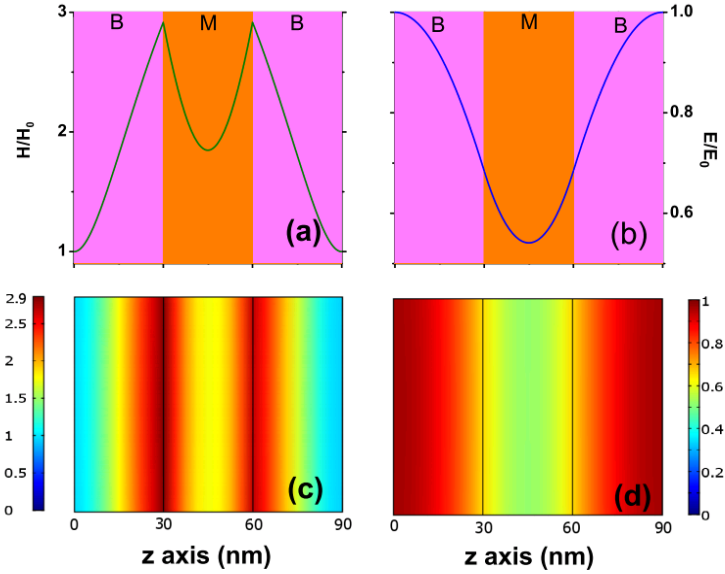

FIG. 4: (Color online) Distribution of local magnetic field enhancement |H/H$_0$| and local electric field enhancement |E/E$_0$| in the birefringent BMB structure under normal incidence. Fig. 4(a) and (b) are theoretical results of EM fields along the z axis; Fig. 4(c) and (d) are simulated results on y-z plane. The other parameters are the same as those in Fig. 1.

We consider a unit amplitude plane wave impinging normally on the birefringent BMB structure. Then the magnetic field and the electric field in the structure can be calculated via TMM [15, 16]. As is shown in Fig. 4(a) and (b), the magnetic field grows rapidly in layer B and reaches its maximum at the first B-M interface, and then after passing the metallic film, it restores the former magnitude at the second B-M interface. The magnitude of the electric field in birefringent layer B undergoes a contrary process. Instead of usual exponentially decaying waves in one single metallic film, the electromagnetic wave restores its phase and amplitude in the BMB structure, which leads to the perfect transmission. In Fig. 4(c) and (d), it reveals that the wave fronts are kept and EM waves are tunneled. The simulated results are in good agreement with our theoretical results.

Next, we consider a more realistic situation, i.e., lossy and frequency-dependent BMB model. The permittivity of the metallic film is assumed to be Drude-type,

$$\varepsilon_M(\omega) = 1 - \frac{\omega_p^2}{\omega^2 + i\gamma\omega} \qquad (4)$$

where $\omega_p$ is the plasma angular frequency and $\gamma$ is the damping coefficient. The loss tangent of the metallic film is chosen to be about $10^{-2}$, and weak dissipation (loss tangent $=10^{-3}$) is assumed for birefringent coatings [19]. The parameters are $\omega_p = \sqrt{30}\omega_0$, $\gamma = 1.765 \times 10^{-3}\omega_p$, i.e., Re[$\varepsilon_M(\omega_0)$] ≈ −29 for the metallic film. We use Eq. (3) and EMT to obtain perfect transmission solutions for the given thickness (d=30 nm) of lossy BMB structure. From Fig. 5(a), we can see that two solutions of Eq. (3) merge together at a critical frequency (about $1.6\omega_0$). Meanwhile, the EMT value is in the middle of them under the critical frequency, obviously implying EMT is no longer accurate. Because the permittivity of the metallic film increases very fast with the decrease of frequency, long-wavelength limit cannot be satisfied in the low frequency. Fig. 5(b) shows the transmittance of a single metallic film and BMB lossy model as the functions of angular frequency at normal incidence. Compared with almost opaque single metallic film, the transmittance of BMB structure is significantly improved around $\omega_0$ and high transmittance is maintained for a large range over frequencies.

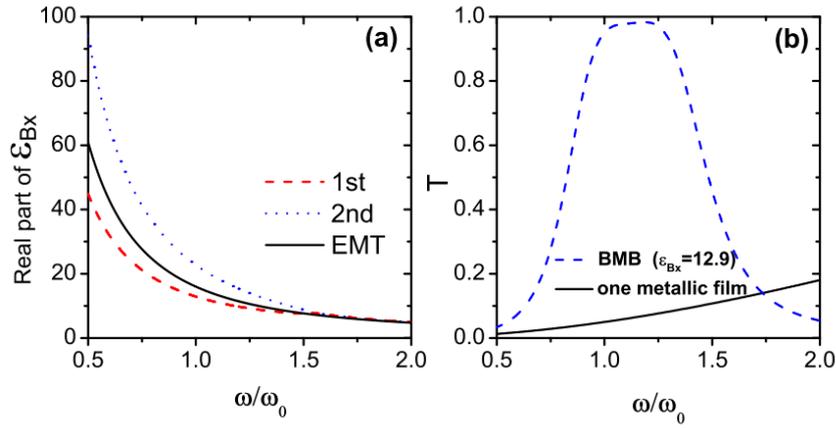

FIG. 5: (Color online) (a)Solutions of Eq. (3) and EMT as a function of the incident frequency for BMB structure with slightly loss. The thickness for every layer is 30 nm. (b)Transmittance as the functions of the angular frequency at normal incidence.

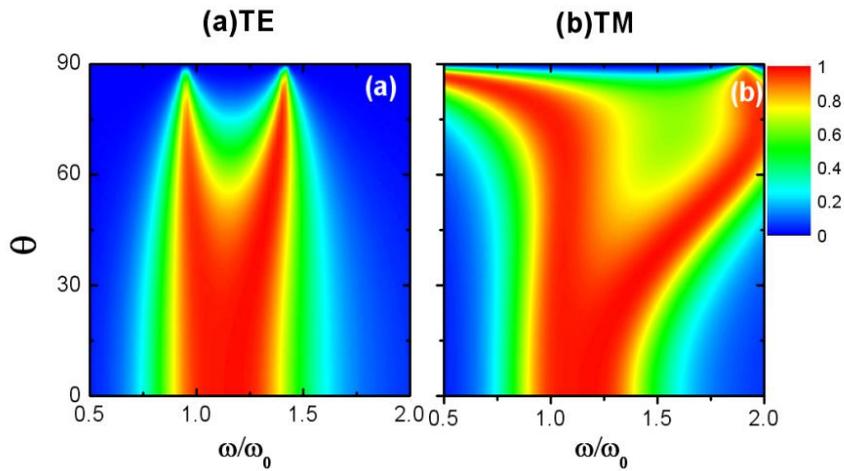

FIG. 6: (Color online) the color: Transmittance as the functions of the angular frequency and incident angle for (a) TE and (b) TM waves. The permittivity of birefringent coatings are $\varepsilon_{Bx}=12.90$ and $\varepsilon_{Bz}=0.56$. The other parameters are the same as those in Fig. 5.

Now, we use Eq. (3) to obtain solutions of the birefringent coatings for the BMB lossy model at the frequency of $\omega_0$. Fig. 6 shows the transmittance as the functions of angular frequency and incident angle for both TE and TM waves. Transparency can be observed in a large range of frequencies (from $0.95\omega_0$ to $1.3\omega_0$) for both TE and TM waves. The transparent angles are not very selective in this frequency range. The broad transparency for higher frequencies owing to the decreasing opacity and loss of the metallic film as well as the permittivity parameters meet the criterion given by Eq. (3). Interestingly, two all-angle transparencies (at the frequencies of $\omega_0$ and $1.3\omega_0$) exist for TE wave, as shown in Fig. 6(a). The first one $\omega_0$ appears where we expected, while the emerge of second one is because $\varepsilon_{Bx}$ of birefringent coatings in this case is just equal to one of the solutions (about 12.90) for the case of $1.3\omega_0$ according to the criterion given by Eq. (3) (see Fig. 5(a)). However, $\varepsilon_{Bz}$ does not satisfy the criterion of the case of $\omega = 1.3\omega_0$, so for TM wave the second one is severely distorted as shown in Fig. 6(b). This phenomenon can be explained. The electric field of TE wave only oscillate in x-y plane, while TM wave also has vector component of the electric field in z-axis direction when waves is incident obliquely into the structure. The propagation of TM waves is related to both $\varepsilon_{Bx}$ and $\varepsilon_{Bz}$, which makes it more difficult to realize all-angle transparency. That's why we introduce birefringent coatings so as to satisfy the transparency criterions for both waves at the same time.

In conclusion, we have shown that metallic films with two identical birefringent layers on both sides can be transparent over almost all directions for both TE and TM polarized waves. The criterion for perfect transmission in the birefringent BMB sandwich structure is obtained, which works quite well for both lossless and lossy cases. EMT results are derived in the long-wavelength limit. Even when the long-wavelength limit is not satisfied, the criterion is still valid. Simulations of various incidences, e.g., point source and plane waves, are in good agreement with our theoretical analysis. The distinguished feature of the proposed BMB structure, i.e., polarization-independent and perfect transmission for a wide range of the angles of incidence, may have potential and practical applications in designing optical and quantum mechanical devices, with much loosened parameter requirements.

**Acknowledgments**

This work was supported by the National Natural Science Foundation of China under Grant No. 11074183, the Key Project in Natural Science Foundation of Jiangsu Education Committee of China under the Grant No. 10KJA140044, the Key Project in Science and Technology Innovation Cultivation Program, the Plan of Dongwu Scholar, Soochow University, and the Project Funded by the Priority Academic Program Development of Jiangsu Higher Education Institutions. We also acknowledge the support of grant R-263-000-574-133 from the National University of Singapore.